\documentstyle[12pt,epsf,epsfig,a4wide]{article}
\begin{document}
\def\beq{\begin{equation}}
\def\eeq{\end{equation}}
\newcommand{\form}[1]{(\ref{#1})}
\begin{centering}
\begin{flushright}
CERN-PH-TH/2005-186, ACT-10-05, MIFP-05-26 \\
October 2005
\end{flushright}
\vspace{0.1in}
{\Large {\bf Robust Limits on Lorentz Violation from Gamma-Ray Bursts}}

\vspace{0.3in}

{\bf John~Ellis}$^{a}$, {\bf N.E.~Mavromatos}$^{b}$,
{\bf D.V.~Nanopoulos}$^{c}$, \\ {\bf A.S.~Sakharov$^{a,d}$} {\bf and} {\bf
E.K.G.~Sarkisyan$^{e}$}
\vspace{0.2in}

\vspace{0.4in}
\end{centering}
\begin{center}
{\bf Abstract}
\end{center}
\vspace{0.2in}

{\small\noindent

We constrain the possibility of a non-trivial refractive index in free
space corresponding to an energy-dependent velocity of light: $c(E) \simeq
c_0 ( 1 - E/M)$, where $M$ is
a mass scale that might represent
effect of quantum-gravitational space-time
foam,
using the arrival times of sharp features observed in the
intensities of radiation with different energies from a large sample of
gamma-ray bursters (GRBs) with known redshifts. We use wavelet techniques
to identify genuine features, which we confirm in simulations with
artificial added noise. Using the weighted averages of the time-lags
calculated using correlated features in all the GRB light curves, we find
a systematic tendency for more energetic photons to arrive earlier.
However, there is a very strong correlation between the parameters
characterizing an intrinsic time-lag at the source and a
distance-dependent propagation effect. Moreover, the significance of the
earlier arrival times is less evident for a subsample of more robust
spectral structures. Allowing for intrinsic stochastic time-lags in these
features, we establish a statistically robust lower limit: $M > 0.9 \times
10^{16}$~GeV on the scale of violation of Lorentz invariance.}

\vspace{0.5in}
\begin{flushleft}
CERN-PH-TH/2005-186 \\
October 2005
\end{flushleft}

\vspace{0.25in}
\begin{flushleft}{\footnotesize
$^a$ Theory Division, Physics Department, CERN, 1211 Geneva 23,
Switzerland \\
$^b$ Theoretical Physics, Department of Physics, King's College London,
Strand, London WC2R 2LS, UK\\
$^c$ Department of Physics, Texas A \& M University, College Station,
TX~77843, USA; \\
Astroparticle Physics Group, Houston Advanced Research Center (HARC),
Mitchell Campus, Woodlands, TX~77381, USA; \\
Academy of Athens,
Division of Natural Sciences, 28~Panepistimiou Avenue, Athens 10679,
Greece \\
$^d$ Swiss Institute of Technology, ETH-Z\"urich, 8093 Z\"urich,
Switzerland \\
$^e$ EP Division, Physics Department, CERN, 1211 Geneva 23,
Switzerland; \\
Department of Physics, the University of Manchester, Manchester M13
9PL, UK  \\}
\end{flushleft}

\newpage
\section{Introduction}

The construction of a quantum theory of gravity remains an elusive
goal~\cite{will}, even more so the formulation of incisive
experimental tests of such a
theory. There is a general expectation that quantum fluctuations in the
background space-time metric would make it appear `foamy' on short time
and distance scales~\cite{foam}, for which models have been proposed in
various frameworks such as non-critical string theory~\cite{EMN}, loop
quantum gravity~\cite{mad}, string theory~\cite{string_kostoletsky},
double special relativity~\cite{dsr} and an effective field theory
approach~\cite{pospelov}. It
has been suggested that the propagation of
light through this space-time foam might exhibit a non-trivial dispersion
relation {\it in vacuo}~\cite{amellis}, corresponding to Lorentz violation
via an energy-dependent velocity of light. It was also pointed out that
one powerful way to probe this possibility may be provided by distant
astrophysical sources of energetic photons that exhibit significant and
rapid variations in time, such as gamma-ray bursters
(GRBs)~\cite{amellis}. The possibility of an energy-dependent velocity of
light:~$c(E) \simeq c_0 ( 1 - E/M)$, where $c_0$ is a limiting low-energy
velocity of light and $M$ is
a mass scale that might represent
effect of quantum-gravitational space-time
foam,
 has subsequently been explored and
constrained by many phenomenological analyses using pulsars and active
galactic nuclei (AGNs) as well as GRBs. Various limits in the range $M >
10^{15}$ to $10^{17}$~GeV have been
reported~\cite{mkr421,onegrb,newonegrb,pulsar,limits}.  The
importance of Lorentz violation for high-energy cosmic rays has also been
considered~\cite{cosmics}, and laboratory probes of different forms of
Lorentz violation have also been investigated~\cite{Kostal}.

In order to identify an effect as radical as Lorentz violation, one must
minimize the uncertainties by a careful statistical analysis and control
of possible systematic errors. One cannot rely on a single source, for
which it would be impossible to distinguish an intrinsic time-lag at the
origin from a delay induced by propagation {\it in vacuo}, particularly if
the observation of the source is uncertain in a crucial energy band. For
this reason, we have pioneered the systematic analysis~\cite{wavegrb} of
statistical samples of GRBs at a range of different redshifts, and we have
introduced techniques from signal processing such as wavelet analysis to
identify and correlate genuine features in the intensities observed in
different energy bands. This technique has the advantage that it can
extract time-dependent features from the signals of many GRBs, even weak
ones, comparing measurements with more than one spectral channel. Our
original analysis~\cite{wavegrb} used data from the BATSE and OSSE
instruments on the
CGRO satellite~\cite{cgro}. Recently, several new instruments such as
HETE~\cite{hete} and SWIFT~\cite{swift} have provided additional, larger
samples of GRBs with known
redshifts.

In this paper, we combine these newer data with the older BATSE data to
establish a hard limit on the foamy mass scale $M$. We supplement the
wavelet technique used earlier with a statistical noise technique for
identifying genuine features in the light curves. We also make a careful
treatment of systematic uncertainties associated with the possibility of
intrinsic time-lags in the sources and treat the
data with standard tools for evaluating the statistical significance of
possible signals. A two-parameter fit to $M$ and a fixed
time-lag would seem to suggest a significant foamy effect. However, the
two parameters are highly correlated and the quality of the fit is very
poor, suggesting the existence of some additional source of error and/or
uncertainty. If one does not know why a fit is poor, the Particle
Data Group
suggests rescaling the estimated errors so that the overall
$\chi^2$/d.o.f. $ \simeq 1$.  Applying this procedure to the full data set
of time-lags in features found in a large sample of GRBs, we are left with
{\it a priori} 4$\sigma$ evidence that higher-energy photons tend to
arrive earlier than lower-energy photons. Selecting the most
robust spectral features reduces the significance of the time-delay effect
to the 2$\sigma$ level when we rescale the errors in this way. However,
one natural hypothesis for the origin for the poorness of the fit is that
the
intrinsic time-lags at the sources are distributed stochastically.
Including this possibility in an analysis of the full data set and
choosing the r.m.s. spread of the intrinsic time-lags so that the the
overall $\chi^2$/d.o.f. $ \simeq 1$, we find that the significance of any
propagation effect is reduced to the 1$\sigma$ level.

Using the robust subsample and either rescaling the errors or allowing for
stochastic intrinsic time-lags, we find rather similar lower limits on the
quantum-gravity scale $M$, which are similar whether we estimate them
within a a Bayesian approach or use the likelihood function. Taking the
smallest of the numerical limits on $M$, we find onservatively $M > 0.9
\times 10^{16}$~GeV at the 95\% confidence level. There is no stronger
robust limit on such a form of Lorentz violation.

\section{Time Delays induced by Lorentz Violation in the Expanding
Universe}

To search for possible effects of violation of Lorentz invariance,
we compare the propagation of photons with energies much
smaller than the mass scale $M$ characterizing the difference of the
vacuum
refractive index from unity~\footnote{In four-dimensional models of
quantum gravity this scale may
be of the same order as the Planck mass $M_P$, but it might be much
smaller in some models with large extra dimensions~\cite{extraellis}.}. A
small
difference between the velocities of two photons with an
energy difference $\Delta E$, emitted simultaneously by a remote
cosmological source, would lead to a time-lag between the arrival
times of the photons.
Taking into account that the effect on the propagation of photons
due to the expansion of the Universe, one has~\cite{wavegrb}
the following
dependence on
redshift $z$
of  the induced differences in the
arrival
times of the two photons with energy difference  $\Delta E$:
\beq
\label{timedel1}
\Delta t_{\rm LV}=H_0^{-1}\frac{\Delta
E}{M}\int\limits_0^z\frac{dz}{h(z)},
\eeq
where $H_0$ is the Hubble expansion rate and
\beq
\label{h}
h(z) = \sqrt{\Omega_{\Lambda} + \Omega_M (1 + z)^3}.
\eeq
Throughout this paper, we assume a spatially-flat
Universe: $\Omega_{\rm total} = \Omega_{\Lambda} + \Omega_M = 1$ with
cosmological constant $\Omega_{\Lambda} \simeq 0.7$:
see~\cite{starobinsky}
and references therein.

To look for such a vacuum
refractive index effect, we need a
distant, transient source of photons of different energies,
preferably as high as possible. One may then measure
the differences in the arrival times of sharp transitions in the
signals in
different energy bands. GRBs are at cosmological distances, as
inferred from their redshifts, and exhibit many transient features in
their time series in different energy bands. In comparison, the
observed active galactic nuclei (AGNs) have lower redshifts and broader
time structures in their emissions, but have the advantage of higher
photon energies. Observable pulsars~\cite{pulsar} have very well-defined
time
structures in their emissions, but are only at galactic distances.

The key issue in all such probes is to distinguish the effects of the
possible violation of Lorenz invariance from any intrinsic delay in the
emission of photons of different energies by the source. It is obvious
from Eq. \form{timedel1} that any effect of violation of Lorentz
invariance should increase with the redshift of the source, whereas source
effects would be independent of the redshift in the absence of any
cosmological evolution effects~\cite{ellisbound}. Therefore, in order to
disentangle source and propagation effects, it is preferable to use
transient sources with a broad spread in known redshifts $z$. Thus, one of
the
most model-independent ways to probe the time-lags that might arise from
quantum gravity is to use the GRBs with known redshifts, which range up to
$z
\sim 6$.

In the present paper, we exploit a sample of 35 GRBs with known
redshifts, including 9 GRBs detected by the Burst And Transient
Source Experiment (BATSE) aboard the Compton Gamma Ray Observatory
(CGRO),
15 detected by the High Energy Transient Explorer (HETE)
satellite and 11
detected by the SWIFT satellite.  We used public archives for
the light
curves obtained with
BATSE~\cite{cgro}
HETE~\cite{hete} and SWIFT~\cite{swift}.
The information on the redshifts has been collected from~\cite{z_info}.
The GRBs in our sample are listed in
Table~\ref{table1} together with their redshifts [24--45] and the 
time-lags  
we extract from their light curves. The BATSE light curves were
recorded with a time resolution of 64~ms in four spectral channels with
boundaries at approximately 25, 55, 115 and 320~KeV.  The HETE time
resolution is rather coarser, namely 164~ms in three energy bands in a
similar energy range. The light curves of SWIFT are also produced with a
resolution of 64~ms in four energy ranges that are almost identical those
of BATSE. In this analysis, we look for spectral time-lags in the light
curves recorded in the $115-320$~KeV energy band relative to those in the
lowest $25-55$~KeV energy band, thereby maximizing the lever arm in photon
energies provided by the available data.  We renormalize the time-lags
obtained from the HETE data with respect to the two above-mentioned BATSE
energy bands, by a factor corresponding to the energy dependence
in~\form{timedel1} and determined by the ratio of the energy differences
$\Delta E$ between the boundaries of the HETE and BATSE spectral channels.
We also renormalized some of the SWIFT data, because we used more
closely-spaced energy bands for some of the SWIFT GRBs.

{\small \begin{table} [h]
\begin{center} \begin{tabular}{|c c c c|}
\hline
GRB&
$z$&$z$ Refs.&
$\Delta t_{\rm total}^{(\rm E_{high}-E_{low})} {\rm (s)}$
\\
\hline
\hline
 &&
BATSE (64~ms)& \\
\hline
970508&
0.835&\cite{97050}&
-0.059$\pm 0.044$\\
\hline
971214&
3.418&\cite{971214}&
-0.098$\pm 0.045$
\\
\hline
 980329&
3.9&\cite{z_info}&
-0.084$\pm 0.036$\\
\hline
980703&
0.966&\cite{980703}&
0.138$\pm 0.053$
\\
\hline
990123&
1.600&\cite{990123}&
-0.155$\pm 0.041$
\\
\hline
990308&
1.2&\cite{990308}&
0.0188$\pm 0.0138$
\\
\hline
990510&
1.619&\cite{990510}&
-0.0017$\pm 0.0143$
\\
\hline
991216&
1.020&\cite{991216}&
-0.0091$\pm 0.0012$
\\
\hline
990506&
1.3060&\cite{990506}&
-0.0503$\pm 0.0075$
\\
\hline
&&
HETE (164~ms)& \\
\hline
010921&
0.45&\cite{010921}&
0.0357$\pm 0.0585$
\\
\hline
020124&
3.198&\cite{020124}&
-0.0046$\pm 0.0455$
\\
\hline
020903&
0.25&\cite{020903}&
-0.0150$\pm 0.0386$
\\
\hline
020813&
1.25&\cite{020813}&
-0.1602$\pm 0.0794$
\\
\hline
020819&
0.41&\cite{020819}&
0.222$\pm 0.145$
\\
\hline
021004&
2.33&\cite{021004}&
-0.0402$\pm 0.1109$
\\
\hline
021211&
1.01&\cite{z_info}&
-0.0202$\pm 0.0639$
\\
\hline
030226&
1.99&\cite{z_info}&
-0.0227$\pm 0.0568$
\\
\hline
030323&
3.372&\cite{030323}&
-0.0148$\pm 0.0570$
\\
\hline
030328&
1.52&\cite{z_info}&
0.00825$\pm 0.07661$
\\
\hline
030329&
0.168&\cite{030329,z_info}&
0.0037$\pm 0.0219$
\\
\hline
030429&
2.66&\cite{030429}&
-0.0123$\pm 0.0965$
\\
\hline
040924&
0.859&\cite{z_info}&
-0.2516$\pm 0.0801$
\\
\hline
041006&
0.716&\cite{z_info}&
0.1179$\pm 0.1228$
\\
\hline
050408&
1.2357&\cite{z_info}&
-0.0562$\pm 0.0989$
\\
\hline
&&
SWIFT (64~ms)& \\
\hline
050319&
3.24&\cite{050319}&
0.0054$\pm 0.0109$
\\
\hline
050401&
2.9&\cite{z_info}&
-0.0135$\pm 0.0285$
\\
\hline
050416&
0.653&\cite{z_info}&
-0.1491$\pm 0.1075$
\\
\hline
050505&
4.3&\cite{z_info}&
-0.0012$\pm 0.0561$
\\
\hline
050525&
0.606&\cite{z_info,grb050525}&
0.1261$\pm 0.0159$
\\
\hline
050603&
2.821&\cite{z_info}&
-0.0032$\pm 0.0047$
\\
\hline
050724&
0.258&\cite{050724}&
0.131$\pm 0.1681$
\\
\hline
050730&
3.968&\cite{050730}&
0.094$\pm 0.1361$
\\
\hline
050820&
2.612&\cite{z_info}&
0.033$\pm 0.0569$
\\
\hline
050904&
6.29&\cite{050904}&
0.004$\pm 0.0852$
\\
\hline
050922&
2.17&\cite{z_info}&
0.0231$\pm 0.0208$
\\
\hline
\end{tabular}\par \vspace{0.3cm}
\end{center}
\caption{ \it
Data on spectral time-lags for GRBs with known redshifts collected by
the BATSE, HETE and SWIFT instruments. }
\label{table1}
\end{table}
\label{sec_prop}}

\section{Detection of Genuine Time-Dependent Features in GRB Light Curves}

We wish to study a possible redshift-dependent time offset between
different energy bands, and so constrain any possible difference in
the propagation speeds of photons of different energies. In order to
measure the times of flight of photons, we need to identify features in
the GRB light curves where the emission rates change very rapidly. The
first step is to introduce criteria specifying the points in the GRB
light curves where the most significant variations take place, and
characterize the degree of variability at each such point. Here we follow
the approach of~\cite{wavegrb}, using the Lipschitz exponents $\alpha_L$
to identify the variation points of the light curves measured in different
energy bands.  The measurability of the Lipschitz exponents provides a
quantitative criterion for identifying genuine variation points belonging
to different energy bands, which may then be correlated. Specifically,
for this analysis we
identify and use the most singular, so-called `genuine' variation points,
at which
the Lipschitz exponents are smaller than $1$ (for details,
see~\cite{wavegrb}).

Following the technique developed in~\cite{wavegrb}, we first use the
wavelet shrinkage procedure~\cite{donoho} based on the thresholding of the
discrete wavelet transform (DWT). This consists of using using the DWT
transform to break the intensity profile down into successive
approximations on different time scales, starting with a relatively coarse
approximation and progressing to successive levels of detail on finer and
finer scales.  We then remove the wavelet coefficients below a specified
threshold value, and diminish the others by the value of this threshold
value. We then reconstruct the intensity profile by the inverse DWT, using
the thresholded wavelet coefficients. This denoising procedure allows us
to separate the structure of the signal from the noise, while retaining
information about the position of true time structures in the signal.
Subsequently, we apply the continuous wavelet transform (CWT) `zoom'
technique to the denoised intensity profiles, which identifies the times
of genuine variation points and estimates their Lipschitz exponents in
each spectral band we utilize. Genuine variational structures found at
similar times in different spectral bands are considered to be
manifestations of one and the same process at the source, if the values of
their Lipschitz exponents are similar to each other. We identify such
variation points in pairs. Since variation points with $\alpha_L$
substantially exceeding $1$ exhibit only smooth transitions of the signal,
and hence do not mark sharp dynamical changes at the source and are more
likely to be spurious, we discard them from the subsequent analysis.
The light curves of most of the GRBs in our dataset exhibit more than one
genuine variation
point with $\alpha_L < 1$. The differences between the
times of the genuine Lipschitz variation points in the higher-energy
spectral band and their counterparts in the lower-energy spectral band are
taken as measures of the possible time delay.

There are several sources of uncertainties in the analysis outlined above,
which may arise from both our procedure for estimating the DWT intensity
profiles and the CWT `zoom' technique for detecting genuine variation
points.  Going beyond~\cite{wavegrb}, here we check the statistical
stability of the procedure used to determine the positions of the genuine
variation points by making simulations in which a Gaussian distributed
noise was added to the original light curves recorded by the instruments,
in each energy band. The amount of the artificial noise was chosen to be
consistent with the thresholding signal-to-noise ratio (SNR) revealed
after the DWT, and with the power of the original signal measured in each
time bin. Following each iteration of the artificial noise, the DWT
thresholding procedure described above~\cite{wavegrb} and the CWT `zoom'
technique were performed repeatedly, until the values of the amplitudes at
the genuine variation points obtained in each iteration and values of
their Lipshitz exponents approached normal distributions. To achieve
reasonable convergence of the distributions of the above-mentioned
quantities to normal distributions, we had typically to make 1000 to 2000
artificial realizations per energy band. This procedure of contaminating
the real data with artificial noise allows us to accumulate
statistics sufficient
to estimate the errors in the determination of the positions
of the genuine variation points~\cite{barlow_book}. The identifications of
pairs of genuine variation points belonging to different spectral bands
but associated with the same emission process at the source are then made
by looking for equality of the mean values of the Lipshitz exponents
within the standard errors of the set of values obtained in the
simulations~\footnote{This procedure also enables us to tag the most
robust time
structures in the light curves, which we discuss later.}.

\section{Time-lags in Emissions from GRBs and the Limit on the Violation
of
Lorentz
Invariance}

We now discuss our procedure for analyzing the possible existence of a
Lorentz-violating contribution~\form{timedel1} to the observed time-lags.
As discussed already in Section~\ref{sec_prop}, this may
be accompanied by {\it a priori} unknown intrinsic
energy-dependent time-lags caused by
unknown poperties of the sources. We take this possibility into
account by fitting the measured time-lags with the inclusion of a term
$b_{\rm sf}$ specified in the rest frame of the source. Therefore, the
resulting observed arrival time delays are fitted by two
contributions:
\beq
\label{lv+rf}
\Delta t_{\rm obs}=\Delta t_{\rm LV}+b_{\rm sf}(1+z),
\eeq
reflecting the possible effects of
Lorentz violation and
intrinsic source effects, respectively.
Rescaling \form{lv+rf} by a factor $(1+z)$, we arrive at a
simple linear fitting function
\beq
\label{lv+rf+scale}
\frac{\Delta t_{\rm obs}}{1+z}=a_{\rm LV}K+b_{\rm sf},
\eeq
where
\beq
\label{K1}
K \equiv \frac{1}{1+z}\int\limits_0^z\frac{dz}{h(z)}
\eeq
is a non-linear function of the redshift which is related to the measure
of
cosmic distance in \form{timedel1}.
The coefficient $a_{\rm LV}=H_0^{-1}\frac{\Delta
E}{M}$ of the slope in $K$ is connected to
the scale of Lorentz violation, whereas the intercept
$b_{\rm sf}$
represents the possible unknown intrinsic time-lag inherited from the
sources. Initially, we assume this
to be universal in the rest frame of every GRB, but later we
investigate the possibility that this varies stochastically.

\begin{figure}[t]
\centerline{\psfig{file=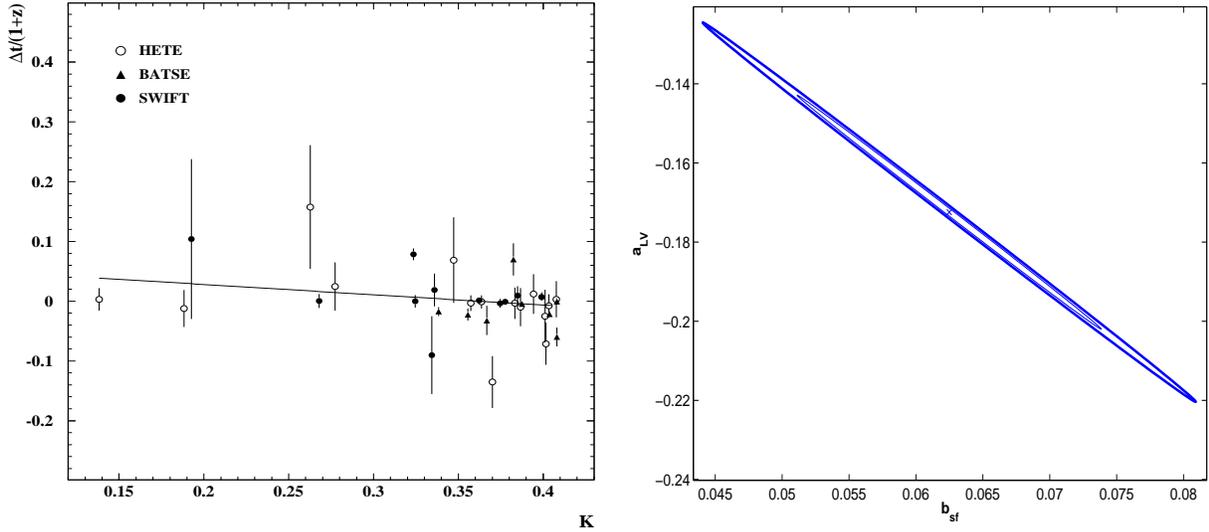,height=7cm,width=16cm}}
\vspace*{8pt}
\caption{
Left panel: {\it
The rescaled
spectral time-lags between the arrival times of pairs of
genuine sharp features
detected in the light curves of the full set of 35 GRBs with
measured redshifts observed by
BATSE (9 light curves with a time resolution of 64~ms),
HETE (15 light curves with a time resolution of 164~ms)
and SWIFT (11 light curves with a time resolution of 64~ms) shown in 
Table~\ref{table1}
normalized to the difference
between the energies of the third and
first BATSE spectral bands. The errors in the redshifts and hence
in $K$ are negligible: the errors in the time-lags are estimated by the
wavelet analysis described in the text.
Also shown in the left panel is a linear fit to these data
with ${\chi^2/{\rm d.o.f.}=148.2/33}$.}
Right panel: {\it
The error ellipse in the slope-intercept
plane for the fit~\form{fir64l}. The 68\% and 95\% 
confidence-level contours are
represented by the inner and outer lines, respectively.}}
\label{regr}
\end{figure}

In order to probe the energy dependence of the velocity of
light that
might be induced by quantum gravity, we compile the
available
data as functions of the variable $K$ (\ref{K1}).
Since \form{lv+rf+scale} exhibits a linear dependence
on $K$, we perform a linear regression analysis of the rescaled
time-lags (\ref{timedel1})
between the genuine variation points in the higher-energy
bands and their Lipschtz
counterparts in the lowest-energy bands of the instruments.
The result
of a straight-line fit~\form{lv+rf+scale} to the rescaled time-lags
extracted from the 35 light curves listed in Table~\ref{table1},
which were collected
by the BATSE, HETE and SWIFT instruments, is shown in the left panel of
Fig.~\ref{regr}~\footnote{As already mentioned, since the
data from the HETE and SWIFT instruments are made available in
slightly different energy bands, we have rescaled the time-lags
and errors from Table~\ref{table1} by the ratios of the energy
differences of the HETE and
SWIFT data relative to those of the BATSE instrument.}.

Although the data points exhibit some scatter, the majority of the
time-lags extracted from the BATSE instrument agree within errors with the
cross-correlation analysis of~\cite{norris2000}. Moreover, our analysis of
the outlying SWIFT detection of GRB050525, whose time-lag we find to be
$0.1261\pm 0.0158$~s, is also in a good agreement with other
calculations~\cite{grb050525}.

It is immediately apparent from Fig.~\ref{regr} that there is a
trend for $\Delta t$ to decrease as $K$ increases. Indeed,
the linear fit shown in the left panel corresponds to
\beq
\label{fir64l}
\frac{\Delta t_{\rm obs}^{\rm tot}}{1+z} \;
= \; (- 0.172 \pm 0.021) \, K\; + \; (0.063 \pm 0.008).
\eeq
The unwary might conclude, prematurely, that there is a very
significant Lorentz-violating effect, since the coefficient
of the linear term in \form{fir64l} is formally many standard
deviations from zero.

However, as seen in the right panel of Fig.~\ref{regr}, the
Lorentz-violating slope $a_{\rm LV}$ and the intrinsic time-lag $b_{\rm 
sf}$ are
very 
highly
correlated. The appropriate tool for estimating the significance of any
Lorentz-violating effect is the marginal distribution of the slope
parameter in the fit~\form{fir64l}, as obtained by integrating over the
intercept parameter. Taking into account the correlation matrix as
described in~\cite{barlow_book}, we rescale by a factor $\sqrt{1-\rho^2}$
the Gaussian-like shape of this marginal distribution, where $\rho$ is the
correlation coefficient of the bivariate slope-intercept distribution. The
mean value is still unchanged at~$a_{\rm LV, \,obs(marg)}^{\rm tot}=-0.172$,
whereas the variance (defined as the width at the half maximum)  is~$\sigma_{a_{\rm LV, \,obs(marg)}^{\rm tot}}=0.018$.
The mean value of the
slope
in this marginalized distribution is therefore negative even with a higher
level of significance. However, the linear fit has a very large value of
$\chi^2/{\rm d.o.f.} = 149/33$, which indicates that either some errors
are underestimated, which is often the case when data from different
instruments are used in a common fit, or there is an additional
uncertainty not yet taken into account.

The first possibility can be modelled by rescaling the errors in
all the extracted time-lags by a scale factor
$S = [\chi^2/d.o.f.]^{1/2}$~\cite{pdg}, guaranteeing that the resulting
$\chi^2/{\rm d.o.f}$ becomes unity. The corresponding linear fit has
central values identical to those
found with the original estimated errors, but the errors in both the tilt
and the intercept are
increased by a factor $\sqrt{S}$:
\beq
\label{fir64l'}
\left(\frac{\Delta t_{\rm obs}^{\rm tot}}{1+z}\right)_{\rm scale} \; = \;
(- 0.172 \pm 0.045) \, K \; + \; (0.062 \pm 0.017).
\eeq
This procedure yields naively 3.8$\sigma$ evidence for a negative
tilt in~\form{fir64l'}, which becomes 4.5$\sigma$ when one calculates
the marginalized distribution for the slope parameter $a_{\rm LV}$. This 
would
still suggest significant evidence for Lorentz
violation, with the null hypothesis
of Lorentz invariance
having a probability of only $10^{-5}$.

\begin{figure}[t]
\centerline{\psfig{file=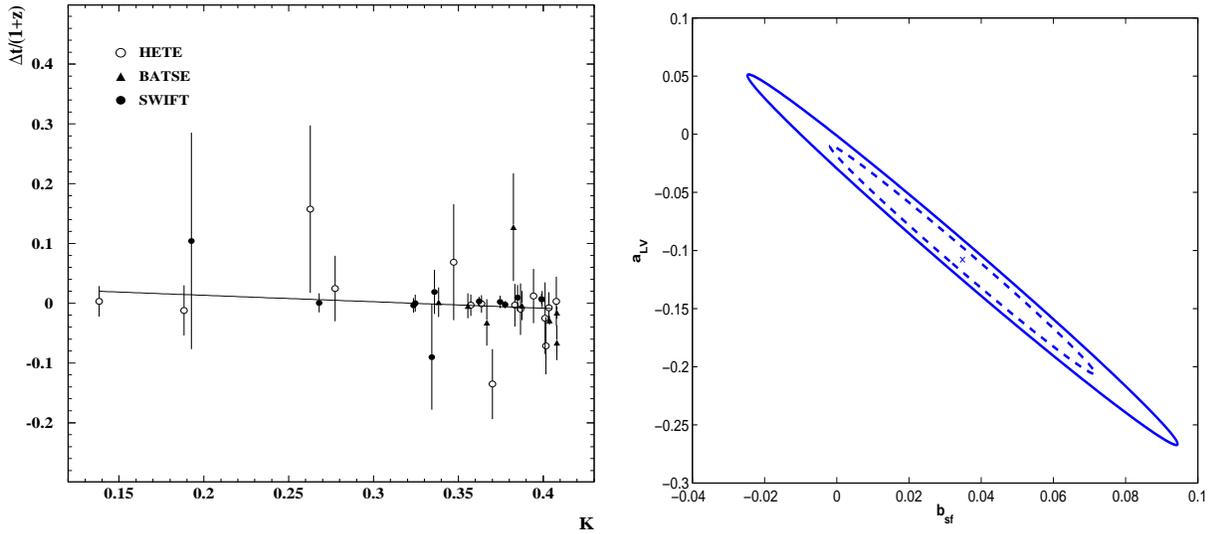,height=7cm,width=16cm}}
\vspace*{8pt}
\caption{
Left panel: {\it
Same as Fig.~\ref{regr} but
using only those high-intensity genuine
variation points thought to be more reliable. Also, the errors have
been rescaled by a universal factor $S$
according to the procedure~\protect\cite{pdg} described in the text, so
that $\chi^2/{\rm
d.o.f.}=1$}.
Right panel: {\it
The error ellipse in the slope-intercept
plane for the fit~\form{fitdw}. The 68\% and 95\% confidence-level 
contours are
represented by the inner and outer lines, respectively.}}
\label{regrfew}
\end{figure}

However, any exceptional claim requires exceptional evidence,
so we must scrutinize this `signal' very carefully.
Instead of rescaling all the errors
uniformly, one might prefer to deweight the time-lags associated with some
of the features whose identification is less reliable.
In the absence of clear understanding of the GRB sources,
we propose the following criterion for selecting a subsample of
the time structures whose analysis may be less unreliable.
Following~\cite{piranphysrep}, we assume that
individual pulses are mostly related to single emission episodes.
Therefore, we retain only those `genuine' variation points which
appear in the
vicinity of the highest pulse in a given light curve,
whereas in the previous overall fit in the left panel of
Fig.~\ref{regr} we used the weighted averages of the differences
between the positions of all the `genuine'
variation points and their counterparts in softer energy bands found in
a given light curve. Specifically, we select only those variation
points at which the
amplitudes reach 70\% of the maximal count rate for a given light curve.
In addition to this choice, we further
restrict our attention to the variation structure with the
highest signal-to-noise ratio in any given light curve.
The resulting selection of time-lags is shown in the left panel of
Fig~\ref{regrfew}, together with a linear fit to the de-weighted data.
Although the data still exhibit significant scatter, the
furthest outlying points have now disappeared.
The corresponding slope and intercept parameters are given by the
following fit, after rescaling by the appropriate $S$ factor:
\beq
\label{fitdw}
\left(\frac{\Delta t_{\rm obs}^{\rm dw}}{1+z}\right)_{\rm scale}
= (-0.108 \pm 0.065) \, K \; + \; (0.035 \pm 0.024).
\eeq
This fit also indicates a negative
value of the slope parameter.
The strong correlation between the slope and intrinsic time-lag
parameters persists, as seen in the right panel of Fig.~\ref{regrfew}, and
the marginalized distribution for the slope yields a 1.9$\sigma$
effect.

Since the significance
of the effect decreases substantially when we select a supposedly more
reliable subsample, we conclude that there is no robust evidence for a
vacuum refractive index.

The rescaling of the errors by a universal factor $S$ was motivated by our
ignorance of the origins of the systematic effects that are clearly
present in our data set. However, since most of the GRBs exhibit more than
one `genuine' feature (in the Lipschitz/wavelet sense described above),
which generally have different time-lags, a combined spectral lag can only
be a crude measure of the time delays within a given burst. For example,
since the burst evolves in time, its internal dynamics may cause different
pulses within a burst to exhibit different intrinsic time-lags. Various
analysis methods are available to examine these
effects~\cite{norris2000,norris2002,ryde1,ryde2}, to which our wavelet
approach~\cite{wavegrb} may be added. However, the degree of evolution of
time-lags is yet to be well quantified, and there may be other
environmental effects not directly related to the evolutionary stage.

\begin{figure}[t]
\centerline{\psfig{file=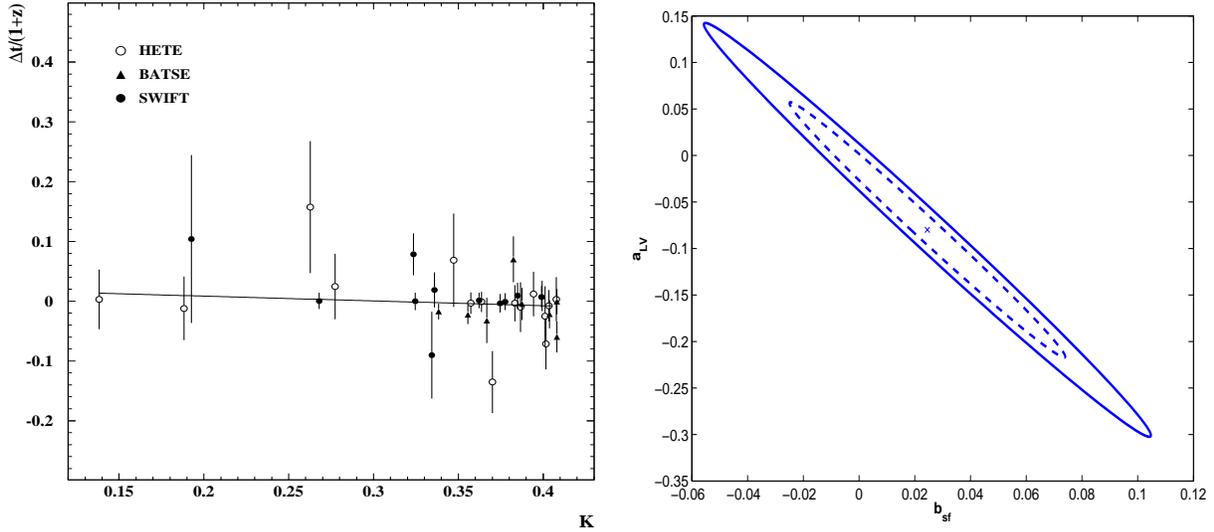,height=7cm,width=16cm}}
\vspace*{8pt}
\caption{Left panel: {\it
Same as Fig.~\ref{regr} but
with the fitted errors in the intrinsic
time-lags increased by 54~ms, modelling a possible stochastic
spread
at the sources, as discussed in the text. The corresponding
linear fit has a $\chi^2/{\rm d.o.f.}$ of unity}.
Right panel: {\it The
error ellipse in the slope-intercept
plane for the fit~\form{fir64l''}. The 68\% and 95\% 
confidence-level contours are
represented by the inner and outer lines, respectively.}}
\label{regr1}
\end{figure}

One simplistic way to take such systematic uncertainties into account
would be, instead of the previous universal rescaling of the errors, to
allow for a universal stochastic spread in the intrinsic time-lag at the
source. This may be done by adding in quadrature, for all the GRBs,
a universal source error whose normalization is then fixed so that
$\chi^2/{\rm d.o.f.}=1$. The corresponding universal source error
is estimated in this way to be 54~ms. A linear-regression fit with this
representation of stochastic intrinsic time-lags is presented in the left 
panel of Fig.~\ref{regr1}, and leads to
\beq
\label{fir64l''}
\left(\frac{\Delta t_{\rm obs}^{\rm tot}}{1+z}\right)_{\rm stochastic}\;
= (-0.080 \pm 0.091)\, K\; +\; (0.025 \pm 0.033).
\eeq
This universal source error of 54~ms is well
inside the resolution of all the instruments whose data we use. The
fit \form{fir64l''} again yields naively a
negative value of the slope parameter, but now only at the 0.9$\sigma$
level. However, the slope and intrinsic time-lag parameters are still
highly correlated, as seen in the right panel of Fig.~\ref{regr1}, and the
significance becomes 1$\sigma$ when the marginalized distribution
is computed. This certainly cannot be considered as evidence for a vacuum
refractive index.

If we restrict our attention to the subset of `more reliable' 
variation points described earlier, and allow for stochastic
fluctuations in their intrinsic time-lags, we find that an error of 
34~ms would
give $\chi^2/{\rm d.o.f.}$ of unity for this subset of variation points. 
This tends to confirm our suggestion that the selected points are indeed 
more reliable. After including this stochastic intrinsic time-lag, the 
linear regression fit for this `reliable' set becomes
\beq
\label{fitdw_stoch}
\left(\frac{\Delta t_{\rm obs}^{\rm dw}}{1+z}\right)_{\rm stochastic}
= (-0.086 \pm 0.074) \, K \; + \; (0.028 \pm 0.027),
\eeq
which is rather similar to \form{fir64l''}.

We therefore quote only a lower limit on the quantum-gravity scale
parameter $M$. To determine this, we first find the regions of the
slope-intercept plane that are allowed at various confidence levels.
In the class of quantum-gravity models we
wish to explore, the slope parameter must be {\it positive}
semi-definite, as also suggested by considering the
possibility of gravitational {\v C}erenkov radiation~\cite{nelson}.
Accordingly, we first quote a
limit on the scale of violation of Lorentz invariance in the
Bayesian manner
proposed
in~\cite{neurtinolimit}, where the confidence range was constructed for
a Gaussian distribution with negative mean which is constrained
physically to
be non-negative. We apply this Bayesian approach to the
`genuine' variation points, with errors rescaled universally: see
(\ref{fitdw}) and Fig.~\ref{regrfew}.
Using the
prescription of~\cite{neurtinolimit} for the measured negative mean of
the marginalized slope distribution at
2$\sigma$ below zero, we calculate the 95\% confidence limit
on
the scale of violation of Lorentz invariance assuming a
random variable obeying Gaussian
statistics with a boundary at the origin.
The corresponding
upper limit on the positive value of the
slope parameter is $a_{\rm LV}^{\rm min}=0.008$. Substituting $a_{\rm
LV}^{\rm min}$
into
the
prefactor formula \form{timedel1}, together with the energy difference
between the upper
and lower BATSE spectral bands, we find the lower limit
\beq
\label{order_stat}
M \ge 1.8 \times 10^{16} \; {\rm GeV}
\eeq
from (\ref{fitdw}) and Fig.~\ref{regrfew}. This becomes
\beq
\label{reliable}
M \ge 0.9 \times 10^{16} \; {\rm GeV}
\eeq
when we use (\ref{fitdw_stoch}). A similar result would be obtained from
(\ref{fir64l''}) and Fig.~\ref{regr1}.

\begin{figure} [t]
\centerline{\psfig{file=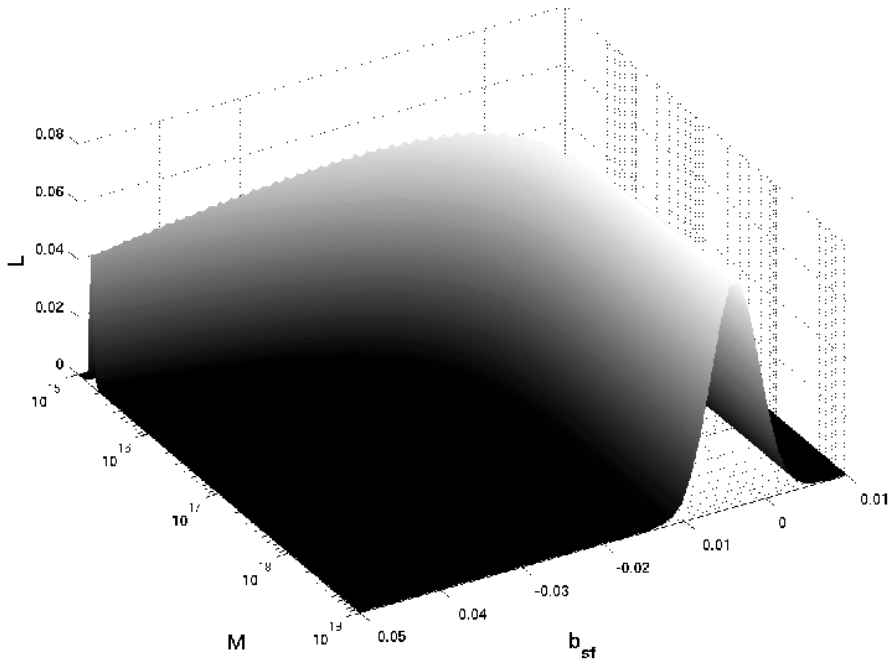,width=12cm}}
\vspace*{8pt}
\caption{\it The likelihood surface in the parameter space of
the intrinsic time-lag $b_{\rm sf}$ and the Lorentz violation scale $M$.}
\label{surf}
\end{figure}

As an alternative to this Bayesian approach, we have also analyzed the
likelihood function $L\propto\exp (-\chi^2(M)/2)$,
where we evaluated $\chi^2$ using
 \beq
\label{chi2}
\chi^2(M)=\sum\limits_{\cal D}\left[\frac{\frac{\Delta
t_i}{1+z_i}-a_{\rm LV}(M)\,K_i-b_{\rm
sf}}{\frac{\sigma_i}{1+z_i}}\right]^2
.
\eeq
The sum in~\form{chi2}
is taken over the all data points $\Delta t_i$, symbolized
by ${\cal D}$, with $\sigma_i$ characterizing the uncertainties in the
measured time-lags.
We have calculated $L$ for all the
possible values of $a_{\rm LV}(M)$, as defined by the coefficient of $K$
in
\form{timedel1} and a reasonable range of the {\it a priori} unknown intercept
parameter $b_{\rm sf}$.
The likelihood surface is shown in Fig.~\ref{surf}.
Marginalizing this likelihood function over the intercept
parameter
$b_{\rm sf}$, one can establish 95\% confidence limit on the scale $M$ of
Lorentz
violation  by
solving the equation
\beq
\label{likl}
\frac{\int_{M}^{M_{\infty}}L_{\rm marg}(\xi
)d\xi}{\int_{0}^{M_{\infty}}L_{\rm marg}(\xi )d\xi}=0.95,
\eeq
where $\infty$  symbolizes a reference
point fixing the normalization. In our case, we choose as reference
point
${M_{\infty}} = 10^{19}$~GeV, the Planck mass.
In practice varying this reference point by as much as an order of
magnitude influences the final result only very weakly.
The 95\% confidence-level lower limit obtained in this way from the more
reliable subset of variation points is
\beq
\label{linlimit}
M\ge 2.1 \times 10^{16}~{\rm GeV}
\eeq
if
we
rescale the errors by a universal factor, and becomes $M\ge 1.6 \times 10^{16}~{\rm GeV}$
if we allow for stochastic
time-lags at the sources for the robust variation points.
It is reassuring that the two ways of calculating the lower
limit on the quantum-gravity scale yield similar numbers
\form{reliable}, \form{linlimit}.
Conservatively, we prefer to quote the weaker lower limit
(\ref{reliable}) at the level of one significant figure.
We have also explored
the effect of dropping any one of the BATSE, SWIFT and HETE
data sets, finding that the tendency of the
slope parameter to be negative is always preserved, though the
significance is (unsurprisingly) reduced somewhat in each case.

\section{Discussion}

The result of the present analysis is significantly stronger and more
robust than that
in~\cite{wavegrb}, thanks to our improved
statistical technique and the use of a
more complete data set. The most conservative limit on the violation of
Lorentz invariance that we find using the most luminous parts of the GRB
emissions is $M \ge 0.9 \times 10^{16}~{\rm GeV}$.
Without this conservative approach, using the whole data set
we would have found strong evidence in favour of Lorentz violation,
with a probability of only $10^{-5}$ for Lorentz invariance.

It is instructive to compare this analysis with that in~\cite{wavegrb},
where we found $M \ge 6.9 \times 10^{15}~{\rm GeV}$.  Here we benefit from
using a larger sample of GRBs with known redshifts, some of which provide
variation points whose time-lags have small errors, and the sensitivity of
this analysis might have been expected to be an order of magnitude greater
than in~\cite{wavegrb}.  However, the strong correlation between the
parameters describing intrinsic and propagation time-lags and the large
value of the overall $\chi^2/{\rm d.o.f.}$ have revealed new systematic
issues. The sensitivity of the new data set to $M$ is reduced by the need
either to rescale the errors~\cite{pdg} or to allow for stochastic
intrinsic time-lags. Moreover, whereas previously we just used a
likelihood analysis, which in this case would have yielded $M \ge 2.1
\times 10^{16}$~GeV (a factor three better than in~\cite{wavegrb}), here
we also use the Bayesian approach~\cite{neurtinolimit}, which yields a
somewhat weaker limit. In order to be conservative, this is the final
limit we wish to quote.

Some stronger upper limits on a linear modification of the photon
dispersion relation have been reported in the literature. These have also
been based on the type of time-of-flight analysis proposed
in~\cite{amellis}, but using individual sources: either a single
GRB~\cite{onegrb,newonegrb} or a single AGN flare~\cite{mkr421}.  As our
analysis shows, any analysis of the time of flight of radiation from a
single source~\cite{mkr421,onegrb,newonegrb} can only be regarded as
indicative, since it is vulnerable to unknown systematic uncertainties
associated with the unknown intrinsic spectral properties of any given
transient source.  In order to establish a rigorous limit, and even more
to establish any non-zero effect, one must be able to distinguish between
intrinsic and propagation effects, which can be done robustly only by
analyzing a sizeable statistical sample of sources. In astronomical
parlance, the quantum-gravity energy-dependent time delay described by
(\ref{timedel1}) is the analogue of a foreground effect superimposed on
the original spectral properties of the GRBs or other sources. The only
way to minimize these unknown intrinsic systematic uncertainties is to
search for a significant linear correlation of the measured time-lags with
the redshifts.  We have learned from our analysis that it is difficult to
break the degeneracy between the slope (propagation) and intercept
(source) parameters.

An analysis~\cite{newonegrb} of the individual
source GRB021206 reported
a stronger lower limit on the scale of
violation of
Lorentz invariance for photon propagation of $M \ge 1.8\times
10^{17}$~GeV, using the relation~\form{timedel1}.
However, the redshift of GRB021206 was not measured directly, and
instead the estimate $z = 0.3 - 0.6$ used in~\cite{newonegrb} was taken
from~\cite{Atteia:2003cb}.
We made the exercise of including GRB021206 in our data fit, assuming
a redshift at either end of this estimated range. If we assume $z=0.3$,
we find $a_{\rm LV}=-0.039\pm 0.016$, which corresponds
to roughly the same
limit as that we obtain. Assuming that this GRB has redshift $z=0.6$
would disfavour a positive slope at the level of 5$\sigma$, but this
is unjustified in the absence of a measured error for the redshift
of GRB021206.
Moreover, assigning to the time-lag for GRB021206 the same
stochastic source uncertainty of 54~ms that we found for our
the other GRBs with known redshifts and combining it with our data set
would give results very
similar to those we quote for either $z=0.3$ or $z=0.6$.

Quite generally, one would need to demonstrate the absence
of any destructive interference between intrinsic and propagation
effects, which has not been done, and
is unlikely to be possible for a single source.
In our view, the only way to disentangle the properties of the source from
propagation effects, and thereby minimize potential systematic errors,
is to look for a correlation with redshift in a statistical sample of GRBs
with a spread of
different measured redshifts. On this basis, the available
GRB data can be used conservatively to set a lower limit
on any possible quantum gravity effect at the level \form{reliable}. The
systematic effects we have found at low and moderate redshifts cast severe
doubts on the use
of individual GRBs with unknown redshifts~\cite{newonegrb,onegrb} to
claim a high
sensitivity to violations of Lorentz
invariance using such a of time-of-flight approach.  

Similar arguments apply to time-of-flight measurements 
using TeV
$\gamma$-flares from blasars~\cite{mkr421}. Moreover, the situation 
with
systematic uncertainties in this case is even less under control than for
GRBs,
because so far
only the single object Mrk421 has exhibited a short-duration
flare, so
there is no way to estimate the systematic
uncertainties inherited from the source,
as we have done here for GRBs using a regression analysis.

Unfortunately, significant progress beyond our limit (\ref{reliable})
may not be possible until the internal dynamics of GRBs is better
understood or data extending to much higher energies become available.

\section*{Acknowledgments}
\noindent
The work of D.V.N. was supported in part
by DOE grant DE-FG03-95-ER-40917.

\end{document}